\documentclass{emulateapj}

\shorttitle{10\micron\ Super-Resolution of T Tau}
\shortauthors{Skemer et al.}

\begin{document}

\title{Evidence for Misaligned Disks in the T Tauri Triple System:\\
10\micron\ Super-Resolution with MMTAO and Markov Chains\altaffilmark{1}}

\author{Andrew J. Skemer, Laird M. Close, Philip M. Hinz, William F. Hoffmann, Matthew A. Kenworthy and Douglas L. Miller}
\affil{Steward Observatory, University of Arizona,
    Tucson, AZ 85721}

\altaffiltext{1}{The observations reported here were obtained at the MMT Observatory, a facility operated jointly by the Smithsonian Institution and the University of Arizona.  Public Access time is available at the MMT Observatory through an agreement with the National Science Foundation.}

\begin{abstract}
Although T Tauri is one of the most studied young objects in astronomy, the nature of its circumstellar environment remains elusive due, in part, to the small angular separation of its three components (North-South and South a-b are separated by 0.68" and 0.12" respectively).  Taking advantage of incredibly stable, high Strehl, PSFs obtained with Mid-IR adaptive optics at the 6.5 meter MMT, we are able to resolve the system on and off the 10 \micron\  silicate dust feature (8.7\micron, 10.55\micron,  and 11.86\micron; 10\% bandwidth), and broad N.  At these wavelengths, South a-b are separated by only $\sim0.3 \lambda/D$. This paper describes a robust Markov Chain Monte Carlo technique to separate all three components astrometrically and photometrically, for the first time, in the mid-IR.  Our results show that the silicate feature previously observed in the unresolved T Tau South binary is dominated by T Tau Sa's absorption, while Sb does not appear to have a significant feature. This suggests that a large circumbinary disk around Sa-Sb is not likely the primary source of cool dust in our line-of-sight, and that T Tau Sa is enshrouded by a nearly edge-on circumstellar disk.  Surprisingly, T Tau Sb does not appear to have a similarly oriented disk.
\end{abstract}

\section{Introduction}
Despite its status as the prototype for young stars, the current perception of T Tauri is that it is an extremely enigmatic system, perhaps abnormally so.  Although T Tauri was originally classified as a single star, \citet{1982ApJ...255L.103D} discovered that it has an infrared companion, which has never been detected in the optical (\citet{1998ApJ...508..736S} place an upper flux limit of $V\sim 19.6$).  \citet{1991AJ....102.2066G} completed an exhaustive speckle-image/slit-scan photometric study to construct SEDs of the strange infrared companion (hereafter T Tau S) along with the original T Tauri (hereafter T Tau N).  The results showed silicate emission in T Tau N and absorption in T Tau S.  The incredibly high infrared luminosity of T Tau S was enough to convince \citet{1991AJ....102.2066G} that it contained its own compact source, and flares in their data indicated the presence of an accretion disk around T Tau S.

\citet{1996AJ....111.2403H} and a followup by \citet{1999ApJ...523..709S} found perpendicular, molecular outflows: an East-West jet is at 23 degrees inclination from the line-of-sight and centered on T Tau N, while a North-South jet is at 79 degrees inclination and centered on T Tau S.  \citet{2002ApJ...568..267K} used integral-field spectroscopy to show that Brackett series emission could be constrained to a small region around T Tau S, which the authors theorized could be indicative of a small edge-on accretion disk.

Compounding the mystery of the infrared companion, \citet{1999aaop.conf..389R} found that T Tau S was non point-like and then \citet{2000ApJ...531L.147K} resolved T Tau S to be a 0.05" binary (hereafter T Tau Sa and T Tau Sb).  When observed again by \citet{2002ApJ...568..771D}, T Tau Sb had moved significantly, implying that Sa is a relatively massive star.  \citet{2002ApJ...568..771D} were also able to resolve Sa from Sb in near-IR spectra, and determined that Sb is a pre-main-sequence, early-type M-star with heavy extinction and active accretion, while Sa's spectrum is generally featureless.

\citet{2004ApJ...614..235B} presented a study of T Tau's IR photometric and spectroscopic variability.  The authors found that T Tau N is not noticeably variable at K or L', while T Tau S varies in the same bands on week-long time-scales.  While changing accretion rates are usually the dominant variability source in classical T Tauri stars, the color variability in T Tau S (K-L' vs K) exhibits a ``redder when faint'' phenomenon, which the authors believe is best explained by variable extinction.  Accretion may also be present, but it cannot explain the color variability by itself.

\citet{2006A&A...457L...9D} were able to use a long baseline of observations to astrometrically determine the masses and orbital properties of the Southern binary, and the results confirmed that T Tau Sa is actually the most massive object in the system (N, Sa and Sb have masses of $\sim$2, $2.73\pm0.31$ and $0.61\pm0.17$ $M_\sun$ respectively).  With some irony, the star that was originally the prototype for the T Tauri star classification, orbits a more massive Herbig Ae star.

Even with the plethora of observations of T Tauri over the last 20 years, the nature of the Southern binary is still a mystery.  Much of this owes to the small separations of the 3 components (N-S and Sa-Sb are separated by 0.68" and 0.12" respectively, using the orbital parameterization from \citet{2006A&A...457L...9D}).  The fact that Sa and Sb are invisible in the optical, and that they can only be split with a powerful AO system limits the temporal and spectral ranges over which the whole system has been studied.  This paper's results, which extend T Tau's resolved photometric range from 4.7\micron\ \citep{2005ApJ...628..832D} to 11.86\micron, will improve our ability to study a resolved SED of the Southern binary. Moreover, the spatial resolution of the silicate feature definitively establishes the stellar source of past unresolved silicate absorption detections.

\section{Observations and Reductions}
We observed T Tauri in 4 filters (8.7\micron, 10.55\micron, 11.86\micron, and N-band) with the 6.5m MMT using the Mid-IR Array Camera, Gen. 4 (MIRAC4), an AO-optimised camera used with the MMT adaptive optics system (MMTAO---see \citet{2000PASP..112..264L} and \citet{2004SPIE.5490...23B}) on Nov. 3, 2006 UT.   MIRAC4 is a super-sampled, 8-25 \micron, 256x256 Si:As array with two optical magnifications.  We used the high-magnification mode (0.055"/pixel) and took 6, 8, 8 and 4 two-minute, chop/nod exposures at 8.7 \micron, 10.55\micron, 11.86\micron (10\% filters), and N-band respectively (Table \ref{Observations}).  We repeated these observations for our PSF-star, Beta Gemini (Pollux).

At the time of the observations, MIRAC4 was in a commissioning, first light run, and had significant electronic artifacts.  Most of the problems are corrected in post-processing using a code written by M. Marengo to remove slowly varying channel biases, cross-talk, echos, and banding (private communication).  Each image is then inspected for residual detector effects (weak pattern noise), and problematic frames are removed (Table \ref{Observations}).  Residual detector artifacts are suppressed by median combining the images.

For mid-IR chop/nod observations at the MMT, the instrument rotator is left off to stabilize background subtraction by always imaging the same warm, reflective surfaces.  So to combine our data, we de-rotate all of our images by the parallactic angle with cubic spline interpolation and cross-correlate the images to align them (standard AO reduction as in \citet{2003ApJ...598L..35C}).  The images are scaled to the maximum image flux of a centered 30x30 pixel range to reduce the effects of atmospheric ozone absorption, and then median combined (see Figure \ref{schematic} for an example of the reduced 10.55\micron\ image we use).

The aligned stack of images is used to construct a sigma image.  In order to robustly calculate the standard deviation of the image stack, the 3 lowest and 3 highest pixels are removed to eliminate outliers (single lowest and highest for broad N).  The sigma image is then multiplied by a constant to correct for this.

Because of the spline shifting, and residual detector effects, we expect the pixel values and their associated errors to be somewhat spatially correlated.  However, this should only cause a slight smoothing effect that will not impact our final results.  We parameterize the image into two error regions---the star region and the background region, and we average the sigma image over the two areas to estimate pixel value errors.

\section{Analysis}
\subsection{Markov Chain PSF Fitting}

In general, it is possible to super-resolve images at angular scales far smaller than $\sim\lambda/D$ given a high enough S/N and PSF-stability \citep{2005ApJ...620..450B}.  A binary-star deconvolution has 6 free parameters: $x_1$, $y_1$, $x_2$, $y_2$, $mag_1$, and $mag_2$.  Adding a third star adds 3 more free parameters. Although, mid-IR AO images have extremely high and stable Strehls, we find that T Tauri is just a little too faint at V-band (MMTAO Strehl starts to drop off slightly for $V>11$ at 10 \micron) to have the near-perfect Strehls our brighter PSF stars achieve with MMTAO \citep{2003ApJ...598L..35C}.  Convolving our PSF stars with a 3-parameter (major-axis, minor-axis, angle) Gaussian ellipsoid is enough to offset this small effect and allow us to work in a mostly photon noise dominated regime.  For the triple system T Tauri, this gives us 6 position parameters, 3 flux parameters, and the 3 PSF modification parameters to fit.

The conventional algorithm to solve a 12-dimensional parameter estimation problem is to do a Levenberg-Marquardt $\chi^2$ minimization \citep{1992nrfa.book.....P}.  Parameter errors may be estimated from the covariance matrix.  However, these error bars are unlikely to be even close to accurate given the complexity of our problem's parameter space.  Constructing a grid of models and calculating relative likelihoods is a much more robust method, but a 12-dimensional parameter space is, in this case, too large to sample with a grid in a reasonable amount of computer time.

Markov Chain Monte Carlo is the widely accepted technique to circumvent the computational unfeasibility of model grids \citep{2002stat.conf,2003bayes.book,2007ApJ...665.1489K}.  Instead of spending significant CPU time calculating model likelihoods in improbable regions, Markov Chains follow a random walk where they spend most of their time in the most likely areas of the parameter space.  The direction of the walk is dictated by the Metropolis-Hastings algorithm, which allows the chain to spend the proper amount of time in each region of parameter space.  Ideally, the Markov chain has converged when the aggregate results no longer have any correlation to the chain's starting position, and the chain has had time to explore every probable part of parameter space multiple times.  Typically, the beginning of the chain is discarded as a ``burn-in" phase.  To be conservative, we always discard the first half of the chain.

Before implementing the Markov chain, we do a Levenberg-Marquardt $\chi^2$ minimization to estimate a reasonable starting value for the Markov Chain.  We also calculate the covariance matrix from the best-fit, and use that to calculate an initial set of jumping conditions.

For each iteration of the Markov chain, we use a covariance matrix to draw a proposed jump from a multinomial Gaussian distribution.  Initially, the covariance matrix is a product of the Levenberg-Marquardt best-fit algorithm, but as the chain progresses, we estimate the covariance matrix from the chain itself.  For each proposed jump, the model's relative likelihood is calculated and compared to the likelihood of the chain's previous step (for Gaussian errors, the relative (unnormalized) likelihood of a model is given by $e^{-\chi^2/2}$).  If the proposed jump is more probable than the previous step, the jump is accepted.  If the proposed jump is less probable than the previous step, it is accepted a certain percentage of the time, equal to the ratio of likelihoods of the proposed jump and the previous step.

\begin{figure}
 \includegraphics[angle=90,width=\columnwidth]{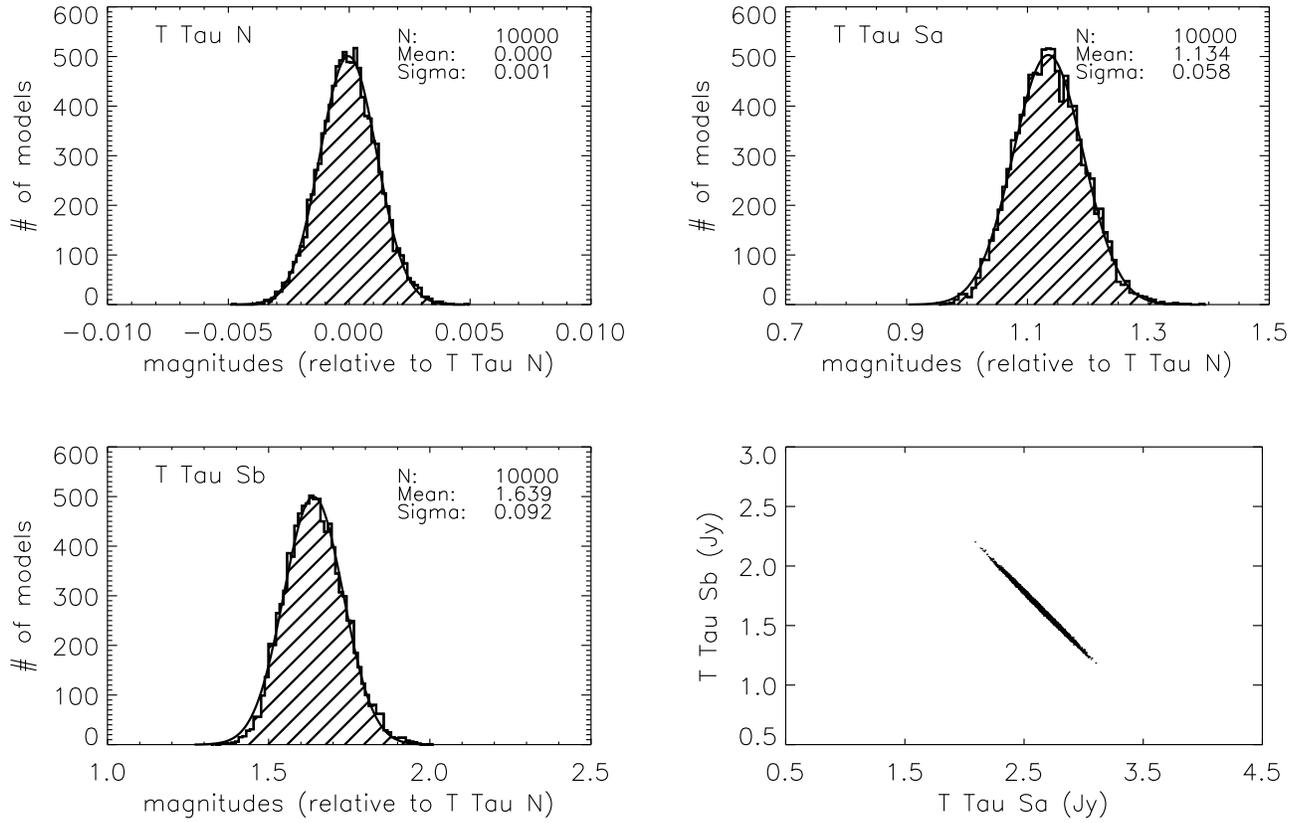}
\caption{Photometric results of the Markov Chains for the 10.55\micron\ filter.  The T Tau N, 
Sa and Sb plots are relative magnitudes scaled to the mean result for T Tau N.  The fourth plot is a scatter plot of fluxes for T Tau Sa vs. T Tau Sb.  The $-1$ slope demonstrates that the T Tau Sa and T Tau Sb flux errors are anticorrelated (if Sa is brighter, Sb is fainter and vice versa).  The fluxes in the fourth plot are relative, and thus ignore the $10\%$ photometric calibration error assumed in the rest of the paper.
\label{mcmc}}
\end{figure}

When the chain has completed enough iterations, the values of each model parameter are binned into a histogram, and fit by a Gaussian.  We show these histograms for the 10.55\micron\ photometric results in Figure \ref{mcmc}.  The smoothness of the histogram for each parameter is a good indication that the Markov Chain has converged.  

At this time, additional quantities of interest may be calculated from the Markov Chain.  For example, we calculate the angular separation of T Tau Sa-Sb using the x-y positions of the components, and use the known 0.68" angular separation of T Tau N and T Tau S (center-of-mass of Sa-Sb, as calculated with the orbital parameterization of \citet{2006A&A...457L...9D}) to measure the platescale.  

Using the Markov Chain Monte Carlo technique for our PSF-fitting problem has provided us with a sophisticated and extremely robust tool to estimate the errors on our relative photometry measurements.  We are also able to quantify the degree to which allowing our PSF to vary slightly is constrained by the data.  The technique is general to almost any deconvolution model, however, it works best with stable, oversampled PSFs, with relatively few free parameters.

\subsection{Convergence of the Chains}
\begin{figure}
 \includegraphics[angle=0,width=\columnwidth]{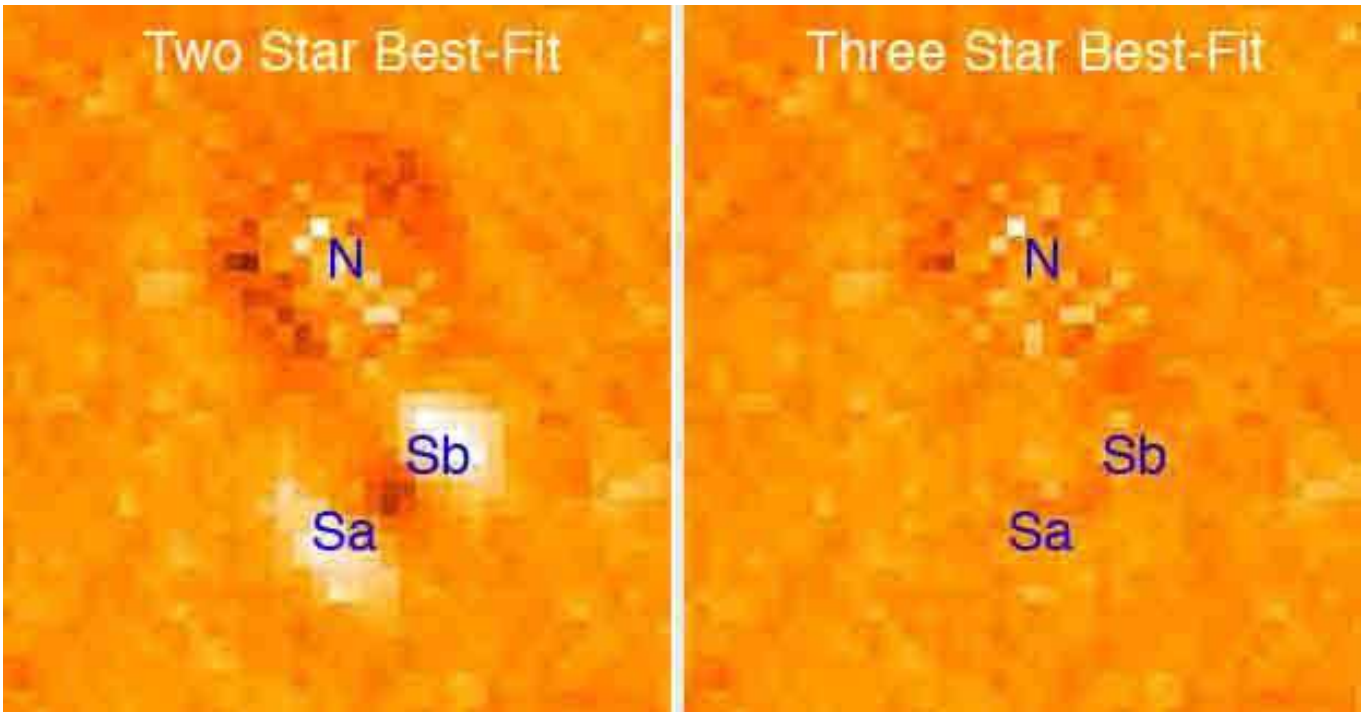}
\caption{Levenberg-Marquardt best-fit residuals at 10.55\micron\ with two-star(left) and three-star models (right).  The three star fit has residuals at the photon
noise floor.
\label{residuals}}
\end{figure}

We verify that the Markov Chains have converged by visually inspecting the histogram of each parameter and confirming that the plots are unimodal and effectively Gaussian (Figure \ref{mcmc}).  We also run our Markov Chain Monte Carlo program multiple times with different starting points and check that our results are unaffected.  A reasonable reduced $\chi^2$ (ie $\sim1$) for our best-fit model is a good indication that our models and error estimations are accurate.  Table \ref{Results} contains the reduced $\chi^2$ of our best-fit models as well as the photometric and astrometric results for each filter.  In Figure \ref{residuals}, we compare the residuals resulting from a two-star (T Tau N; T Tau S) best-fit subtraction, and a three-star (T Tau N; T Tau Sa; T Tau Sb) best-fit subtraction.  The three-star fit has negligible spatial correlations and noise amplitudes consistent with photon noise.

In order to check for systematic effects of our algorithm, we run blind-recovery tests by constructing fake data sets from our resultant parameters and images of Beta Gem (our PSF).  We add noise to the fake data sets consistent with the noise in our real data.  The Markov Chain technique is always able to return the correct parameters within the modeled uncertainties, which conclusively demonstrates that any systematic effects are limited to our slight residual PSF mismatch, and not to any algorithmic bias.

\subsection{Photometry and Astrometry}
We do an absolute photometric calibration based on known fluxes of our PSF star, Beta Gem \citep{mirac3.man}.  Because of $\lesssim 10\%$ varying atmospheric background/absorption during our exposures, we assume an absolute photometric calibration uncertainty of 10\% for each filter, which is a correlated quantity between T Tau N, Sa and Sb.  The Markov Chain technique returns relative errors for each component of T Tau.  T Tau N's relative photometric errors are dwarfed by the absolute calibration uncertainty, while T Tau Sa and Sb's relative errors have an appreciable contribution to their overall photometric uncertainty.  The relative errors on T Tau Sa and Sb are almost completely anticorrelated (Figure \ref{mcmc}d), in the sense that if Sa is brighter, Sb is fainter and vice versa. 

\begin{figure}
 \includegraphics[angle=0,width=\columnwidth]{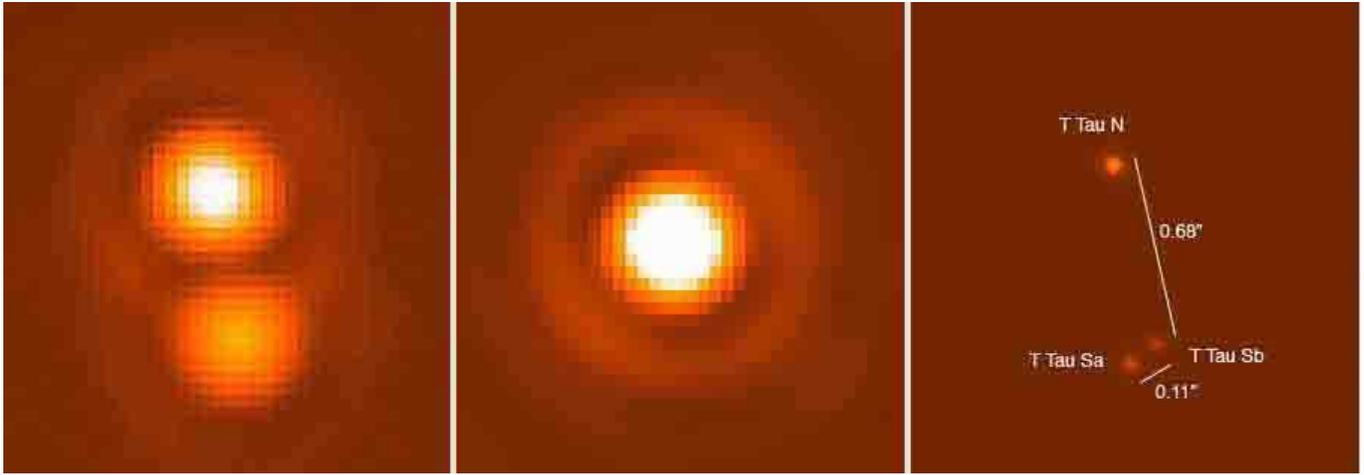}
\caption{For all three graphics, North is up and East is left.  The images are 1.7" on a side and have a $log_{10}$ stretch.\newline
\emph{Left:} The 10.55\micron\ median combined image of T Tauri showing T Tau N and T Tau S.  Here, T Tau S appears to be unresolved.\newline
\emph{Center:} The 10.55\micron\ median combined image of Beta Gemini (the PSF used in this paper).\newline
\emph{Right:} A schematic of the T Tauri system, with 10.55\micron\ photometry and astrometry as derived
in this paper (the N-S separation is from \citet{2006A&A...457L...9D}).  The PSF used in the schematic is scaled down for aesthetics.
\label{schematic}}
\end{figure}

Astrometric quantities should be invariant across the different bandpasses.  We use the orbital parameterization from \citet{2006A&A...457L...9D} along with our measured separation of T Tau N and T Tau S (center-of-mass\footnote{The center-of-mass is calculated by taking our measured separation of T Tau Sa-Sb and assuming a mass ratio for T Tau Sa-Sb as determined by \citet{2006A&A...457L...9D}.}) to determine the plate scale for each observation.  This allows us to measure an angular separation of Sa-Sb, which should be constant in each filter.  We find that our 10.55 \micron, 11.86 \micron\ and N-band calculations are consistent at the 0.01" level ($0.112"\pm0.003"$, $0.104"\pm0.004"$ and $0.121"\pm0.002"$ respectively), and that our 8.7 \micron\ measurement is significantly off ($0.142"\pm0.004"$).  However, poorer seeing conditions during the 8.7 \micron\ observations lead us to believe the longer wavelength results.  Variations in seeing conditions/Strehl can propagate into the strength of the first Airy ring, and because T Tau Sa-Sb is aligned with T Tau N's first Airy ring (see Figure \ref{schematic}a and \ref{schematic}c), separation is a parameter that could be affected by a small, but noticeable, amount.  Systematics in our N-band data (see description below) also lend some doubt to the accuracy of the broad-band results, which could explain why our error bars appear to be slightly underestimated. The \citet{2006A&A...457L...9D} orbital parameters predict a separation of $0.119"\pm0.004"$ at the time of our observations, which is consistent with our measurements at the 0.01" level.

The position angle of the Southern binary is well constrained and consistent across the filter set ($307.4\pm1.1$, $306.5\pm1.6$, $308.4\pm1.1$  and $311.1\pm0.9$ degrees at 8.7\micron, 10.55\micron, 11.86\micron\ and N-band respectively).  The \citet{2006A&A...457L...9D} orbital parameters predict a position angle of $306\pm13$ degrees at the time of our observations, which is consistent with our measurements, although our data place a much tighter constraint.

\begin{figure}
 \includegraphics[angle=0,width=\columnwidth]{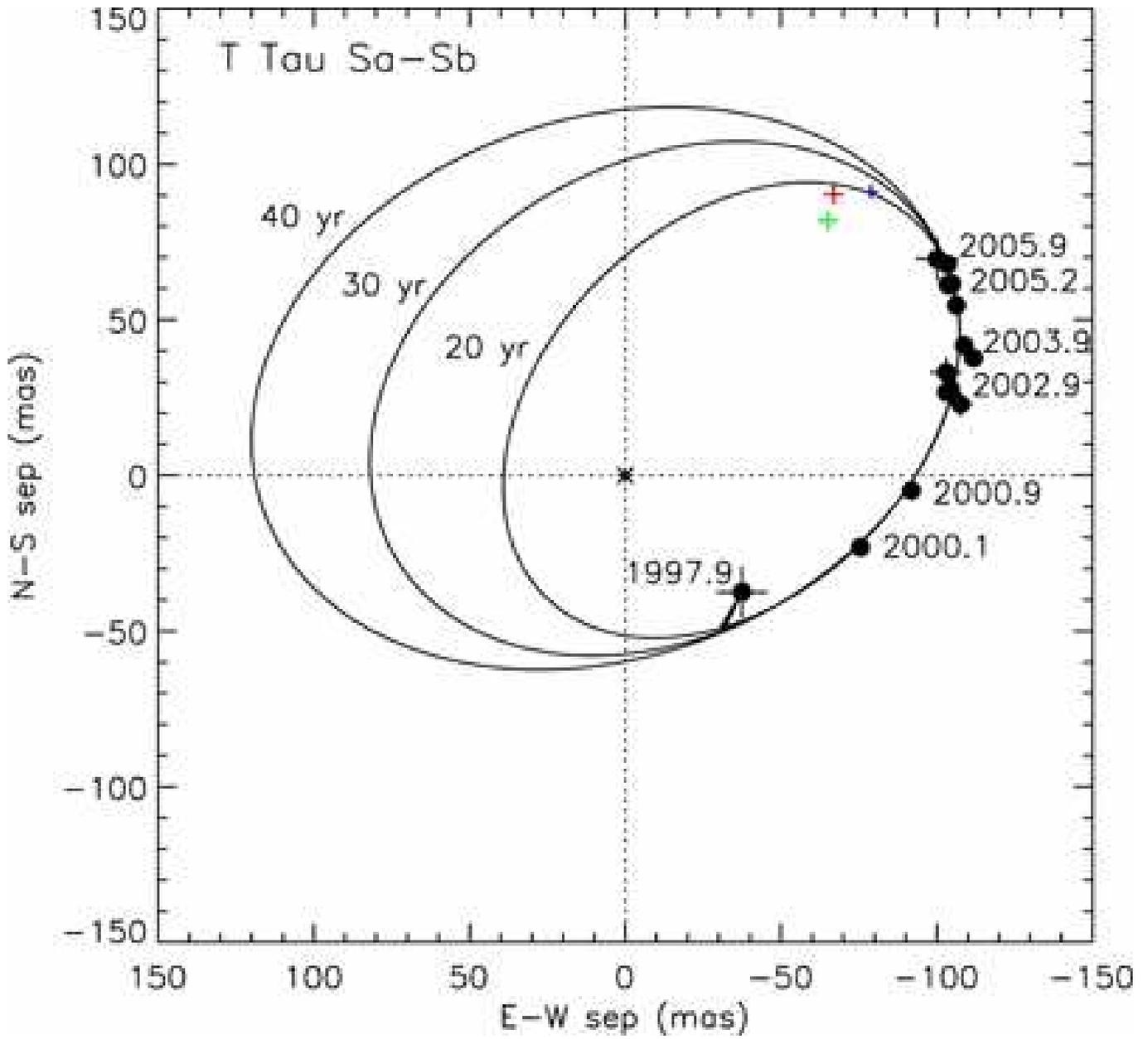}
\caption{Figure modified from \citet{2006AJ....132.2618S} \newline
Astrometric solutions to the T Tau Sa-Sb binary (Sa is in the center) with this paper's results (2006.9) for 10.55\micron\ (red), 11.86\micron\ (green), N-band (blue) overplotted.  There is some preference for orbits with a period shorter than 40 years.
\label{astrometry}}
\end{figure}

In Figure \ref{astrometry} our astrometry results are compared to the orbital motion measurements of \citet{2006AJ....132.2618S}.  Our results are generally consistent with the predicted orbits.  While there is some preference for orbits shorter than 40 years, our results are meant more as a demonstration of fitting accuracy than as an orbital constraint.

Our N-band results are somewhat questionable because we only had data from two different chop-nod sets (and only six usable images total).  This means we could not adequately suppress the detector's correlated noise by median combining the images, and as a consequence, systematics may have dominated our results at N-band more than for the other filters.  Our best-fit reduced $\chi^{2}$ at N-band is 4.43 (as opposed to 1.24, 0.82 and 1.37 at 8.7\micron, 10.55\micron\ and 11.86\micron\ respectively).

\section{Discussion}
\begin{figure}
 \includegraphics[angle=90,width=\columnwidth]{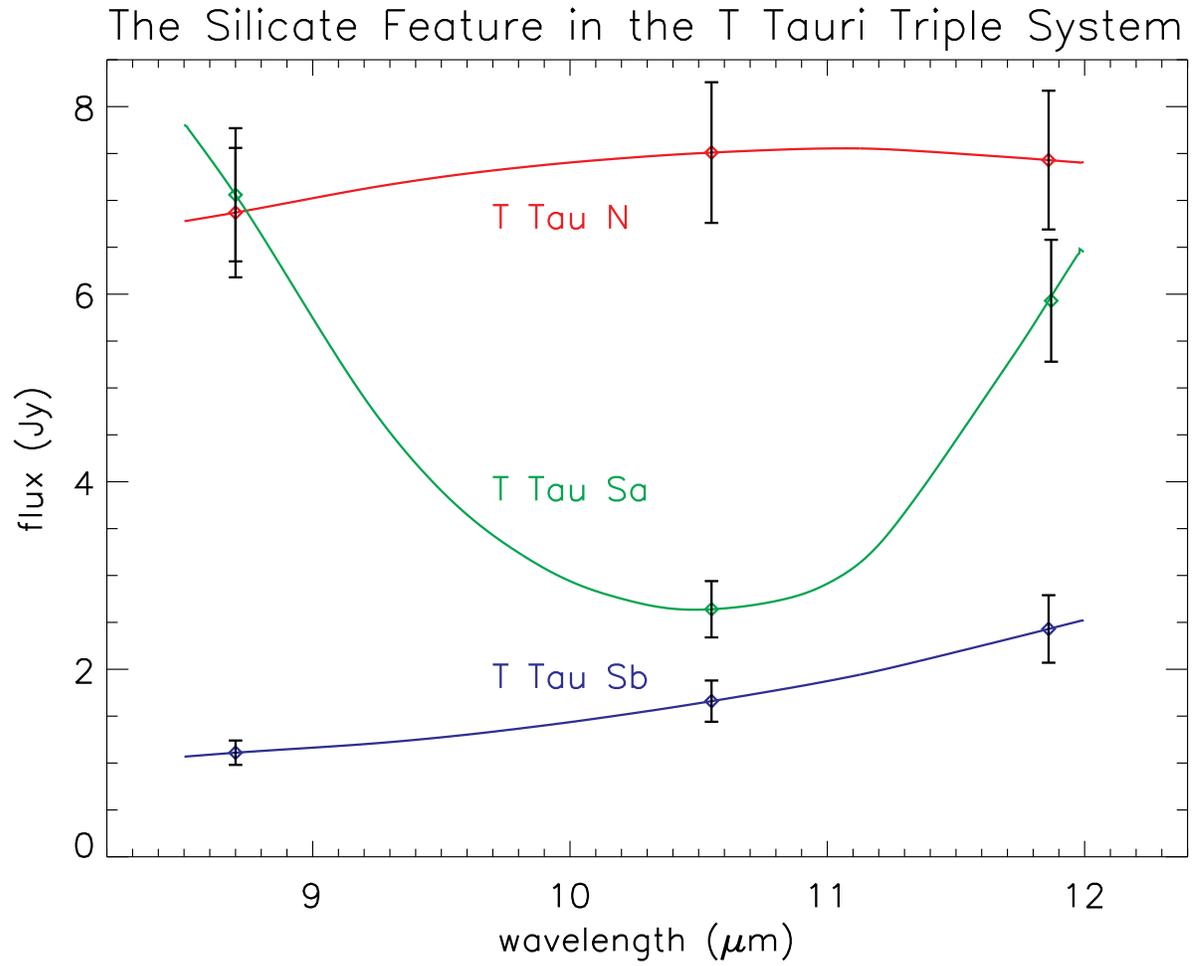}
\caption{The figure shows photometry for 8.7\micron\, 10.55\micron\ and 11.86\micron\ for T Tau N, Sa, and Sb.  The 
curves drawn through the points are intended as a visual aid.  T Tau Sa has a large absorption feature that is absent from 
the other stars, which indicates the presence of an edge-on protoplanetary disk.  Since the other stars lack a similar feature,
it is likely that the disks in the T Tauri system are misaligned.
\label{silicate}}
\end{figure}

Figure \ref{silicate} shows the fluxes and overall flux errors of T Tau N, Sa and Sb in the three narrowband filters.  These results show that the silicate absorption in T Tau S, originally observed by \citet{1991AJ....102.2066G}, originates entirely from T Tau Sa.  T Tau Sb's silicate SED has large enough photometric uncertainty that we cannot say whether it has a small emission or absorption feature.  However, we can say with certainty, that it does not have as dramatic a silicate feature as T Tau Sa.  T Tau N also has a negligible silicate feature.\footnote{
\citet{1991AJ....102.2066G} were able to see an emission feature from T Tau N, but their SED shows that the peak of the emission is probably fairly localized to 9.7\micron, which would explain our null result at 10.55\micron.  The amorphous silicate feature generally peaks at 9.7\micron, however atmospheric ozone absorption makes this wavelength hard to observe from the ground.  Strong features and crystalline silicate can still be detected in the 10.55\micron\ filter.  Our N-band results (using $N-[8.7\micron+10.55\micron+11.86\micron]$ as a proxy for a ``9.7\micron" filter) corroborate the fact that emission/absorption are peaked at 9.7\micron\ (although, as was mentioned in the previous section, the accuracy of the N-band results is questionable). Sb has a higher flux at N-band than any of the narrow-band filters, indicating that Sb could have significant emission at 9.7 \micron.  North also has a reasonably strong N-band point, which suggests that it also probably has silicate emission.} 

\citet{2003AJ....126.3076W} used STIS spectra to infer the presence of circumbinary structure obscuring both objects of the T Tau S binary.  However, a circumbinary structure cannot be the major source of silicate absorption since the absorption is only towards Sa.  Our observations indicate the presence of an edge-on circumstellar disk around T Tau Sa. The dense, optically thick midplane of a disk can obscure the star (causing the high extinction necessary to completely hide a $\sim2.7$ solar mass star shortward of H-band) and create the silicate absorption we observe in an otherwise featureless spectrum.  T Tau Sb is less red, less obscured and has less (if any) silicate absorption than T Tau Sa.  \citet{2002ApJ...568..771D} spectra indicate that T Tau Sb is an M0 star with heavy extinction and active accretion, implying the presence of a disk.  Combined with our null detection of silicate absorption, it is likely that T Tau Sb has a non-edge-on disk.

In general, circumstellar disks in tight binaries are tidally aligned on short time-scales, but \citet{2004ApJ...600..789J} and \citet{2006A&A...446..201M} have found that systems with three or more stars tend to have misaligned disks.  These previous works used polarimetry, and could only separate binaries with $sep > 100 AU$.  By resolving silicate features with 10 \micron\ AO and super-resolution techniques, we can push this limit to $\sim15 AU$ where tidal forces are stronger and disks are truncated.

\section{Conclusions}
Markov chain super-resolution is a useful technique for taking full advantage of highly stable, diffraction limited images from ground-based AO systems or space-based telescopes.  The MMTAO system, along with MIRAC4's supersampled detector, is uniquely capable of producing these images in the mid-IR at a 6.5-meter class telescope.

We have used MMTAO to image the famous T Tauri triple system and were able to split the 0.11" Southern binary on and off the silicate feature (8.7\micron, 10.55\micron, 11.86\micron\ and broad N-band).  While previous unresolved data from the Southern binary show strong silicate absorption (explaining its high extinction), we have determined the source of the absorption is entirely in front of T Tau Sa, a $\sim$2.7 solar mass Herbig Ae star with an otherwise featureless spectrum.

Our results indicate the presence of an edge-on circumstellar disk around Sa, corroborating previous theories resulting from jet orientation, differential extinction of Sa and Sb, narrowly constrained Brackett series emission, and warm, narrow CO absorption \citep{1996AJ....111.2403H,1999ApJ...523..709S,2002ApJ...568..267K,2005ApJ...628..832D}.

T Tau N has been observed to have silicate emission \citep{1991AJ....102.2066G} and has a perpendicular jet orientation to the Southern jet \citep{1996AJ....111.2403H,1999ApJ...523..709S}, which indicates the presence of a face-on disk.  Sb has been modeled with an accretion disk and the lack of silicate absorption in our results means that it also has a non-edge-on disk.  The fact that disks are misaligned in such a tight ($P\sim 20$ year for Sa-Sb) system is surprising considering that tidal forces should align them on short timescales.  As usual, the prototype for young stars has provided another surprise for star formation models.  It remains to be seen whether other tight binaries/triple systems have similar anomalies.

\acknowledgements 
We are grateful to the anonymous referee for his/her extremely careful reading, and insightful report that led to an improved paper.  We particularly want to thank Tracy Beck for useful conversations about the circumstellar environment of T Tauri, in all of its complexity.  We also thank Massimo Marengo for helping to characterize and correct MIRAC4 artifacts.  Brandon Kelly supplied excellent advice and references for constructing a Markov Chain simulation.  Eric Nielsen provided us IDL code to solve binary orbits.  We also thank the MMT staff, in particular Thomas Stalcup, who operated the MMTAO system for the observations in this paper.  AJS and LMC acknowledge NASA Origins and LMC thanks the NSF CAREER program for their generous support.

\bibliographystyle{apj}
\bibliography{database}

\clearpage
\begin{deluxetable}{llcccccccc}
\tabletypesize{\scriptsize}
\tablecaption{Observations of T Tau (Nov. 3, 2006)}
\tablewidth{0pt}
\tablehead{
\colhead{parameter} &
\colhead{8.7\micron} &
\colhead{10.55\micron} &
\colhead{11.86\micron} &
\colhead{N} &
}
\startdata

Chop-Nod Sets\tablenotemark{a} & 6 & 8 & 8 & 4\\
On-Source Time (sec) & 720 & 960 & 960 & 480\\
Usable Images & 20 & 25 & 20 & 6\\
Usable On-Source Time (sec) & 600 & 750 & 600 & 180\\
\enddata
\tablenotetext{a}{Each chop-nod set comprises four images with 8" chops and 6" nods}
\label{Observations}
\end{deluxetable}

\begin{deluxetable}{llllllllll}
\tabletypesize{\scriptsize}
\tablecaption{Photometric and Astrometric Measurements (Nov. 3, 2006)}
\tablewidth{0pt}
\tablehead{
\colhead{parameter} &
\colhead{8.7\micron\tablenotemark{a}} &
\colhead{10.55\micron\tablenotemark{a}} &
\colhead{11.86\micron\tablenotemark{a}} &
\colhead{N\tablenotemark{b}} &
}
\startdata

T Tau N (relative mag)  & $0.000\pm0.002$  & $0.000\pm0.001$ & $0.000\pm0.001$ & $0.000\pm0.002$\\
T Tau Sa (relative mag) & $-0.029\pm0.011$ & $1.134\pm0.058$ & $0.244\pm0.047$ & $0.634\pm0.030$\\
T Tau Sb (relative mag) & $1.977\pm0.075$  & $1.639\pm0.092$ & $1.215\pm0.117$ & $0.702\pm0.033$\\
T Tau Sb-Sa (delta mag) & $2.005\pm0.087$  & $0.498\pm0.151$ & $0.966\pm0.165$ & $0.064\pm0.063$\\
T Tau N (Jy)  & $6.87\pm0.69$ & $7.51\pm0.75$ & $7.43\pm0.74$ & $7.40\pm0.74$\\
T Tau S (Jy)  & $8.17\pm0.82$ & $4.30\pm0.43$ & $8.36\pm0.84$ & $8.01\pm0.80$\\
T Tau Sa (Jy) & $7.06\pm0.71$ & $2.64\pm0.30$ & $5.93\pm0.65$ & $4.13\pm0.41$\\
T Tau Sb (Jy) & $1.11\pm0.13$ & $1.66\pm0.22$ & $2.43\pm0.36$ & $3.88\pm0.39$\\
Sa-Sb Separation (arcsec) & $0.142\pm0.004$ & $0.112\pm0.003$ & $0.104\pm0.004$ & $0.121\pm0.002$\\
Sa-Sb PA (degrees) & $307.4\pm1.1$ & $306.5\pm1.6$ & $308.4\pm1.1$ & $311.1\pm0.9$\\
best reduced $\chi^2$ & $1.24$ & $0.82$ & $1.37$ & $4.43$\\
\enddata
\tablenotetext{a}{10\% filters}
\tablenotetext{b}{8-13 \micron}
\label{Results}
\end{deluxetable}

\clearpage

\clearpage

\clearpage

\clearpage

\clearpage

\clearpage

\end{document}